# Study of Ni-doping Effect of Specific Heat and Transport Properties for $LaFe_{1-y}Ni_yAsO_{0.89}F_{0.11}$

Takayuki Kawamata[1,3#], Erika Satomi[1,3], Yoshiaki Kobayashi[1,3], Masayuki Itoh[1,3] and Masatoshi Sato[1,2,3*,†]

[1] *Department of Physics, Division of Material Science, Nagoya University, Furo-cho, Chikusa-ku, Nagoya 464-8602, Japan*

[2] *Toyota Physical & Chemical Research Institute, Nagakute, Aichi 480-1192, Japan*

[3] *JST, TRIP, Nagoya University, Furo-cho, Chikusa-ku, Nagoya 464-8602, Japan*

Specific heats and transport quantities of the $LaFe_{1-y}Ni_yAsO_{0.89}F_{0.11}$ system have been measured, and the results are discussed together with those reported previously by our group mainly for $LaFe_{1-y}Co_yAsO_{0.89}F_{0.11}$ and $LaFeAsO_{0.89-x}F_{0.11+x}$ systems. The $y$ dependence of the electronic specific heat coefficient $\gamma$ can basically be understood by using the rigid-band picture, where Ni ions provide 2 electrons to the host conduction bands and behave as nonmagnetic impurities. The superconducting transition temperature $T_c$ of $LaFe_{1-y}Ni_yAsO_{0.89}F_{0.11}$ becomes zero, as the carrier density $p$ (=2$y$+0.11) doped to LaFeAsO reaches its critical value $p_c$ ~0.2. This $p_c$ value of ~0.2 is commonly observed for $LaFe_{1-y}Co_yAsO_{0.89}F_{0.11}$ and $LaFeAsO_{0.89-x}F_{0.11+x}$ systems, in which the relations $p = x+0.11$ and $p = y+0.11$ hold, respectively. As we pointed out previously, the critical value corresponds to the disappearance of the hole-Fermi surface. These results indicate that the carrier number solely determines the $T_c$ value. We have not observed appreciable effects of pair breaking, which originates from the nonmagnetic impurity scattering of conduction electrons and strongly suppresses $T_c$ values of systems with sign-reversing of the order parameter over the Fermi surface(s). On the basis of the results, the so-called $s_\pm$ symmetry of the order parameter with the sign-reversing is excluded.

## 1. Introduction

For newly found Fe pnictide superconductors with rather high transition temperature $T_c$,[1)] there has been considerable interest in the symmetry of the superconducting order parameter $\Delta$, because the symmetry gives us important information on the pairing mechanism. If the spin-fluctuation is relevant to the pairing, we expect the so-called the $s_\pm$ symmetry,[2, 3)] which has a sign-reversing between $\Delta$'s on two kinds of Fermi surfaces around $\Gamma$ and M points in the reciprocal space. However, as we have been pointing out from the early stage of the studies,[4-8)] the rates of the $T_c$ suppression observed for Co- and Ru-doped *Ln*1111 systems $LnFe_{1-y}M_yAsO_{0.89}F_{0.11}$ (*Ln* = La and Nd; M=Co and Ru) are too small to understand as the pair-breaking effect for systems with the $s_\pm$ symmetry originating from the conduction electron scattering by nonmagnetic impurities. Because this result is a strong indication that the systems do not have the $s_\pm$ symmetry, we may have to consider a new pairing mechanism involving the orbital degrees of freedom, for example.[9, 10)]

In addition to the above finding, our studies[4-6, 8)] have also revealed that for $LaFe_{1-y}Co_yAsO_{0.89}F_{0.11}$ and $LaFeAsO_{0.89-x}F_{0.11+x}$, the superconductivity disappears at a common critical values of $x$ and $y$. Then, as long as the band shape is not severely changed by the doping, we can expect that the critical value ($p_c$) of the electron numbers $p$ [=($y$+0.11) and ($x$+0.11) for $LaFe_{1-y}Co_yAsO_{0.89}F_{0.11}$ and $LaFeAsO_{0.89-x}F_{0.11+x}$, respectively] is considered to be common, too, and correspond to the electron number at which the doped electrons (almost) completely bury the hole Fermi surfaces around the $\Gamma$ point in the reciprocal space.[5, 6)] The result is important, because it implies that the carrier number solely determine the $T_c$ values and that there is no appreciable effect of the pair breaking which originates, in systems with the sign reversing $\Delta$'s (around $\Gamma$ and M points in the case of the $s_\pm$ symmetry), from the conduction electron scattering by nonmagnetic impurities and brings about the significant difference of the $T_c$-suppression rates or the $p_c$ values between the Co-doped and F-doped systems with and without the scattering centers in the conducting FeAs planes, respectively. Similar behaviors can be also found in the data reported for the $BaFe_2As_2$ system doped with various species of transition metal atoms to Fe sites.[11, 12)]

Here, we measured specific heats and transport quantities of the Ni-doped La1111 system to obtain further information on the impurity effects. First, we have examined the validity of the rigid band picture for Ni-doped La1111 to establish the followings. (i) A doped Ni atom donates two electrons to the host bands and do not carry localized magnetic moment. (Co and Ru atoms donate one and zero electron, respectively.[6)]) (ii) The electronic band structure is not so severely affected by the Ni impurities doped to the conducting planes. (It is also true for the Co doped system,[6)] while for the Ru doped system, the band width becomes larger with increasing $y$ in the relatively large $y$ region.[7, 8)])

If this condition is satisfied, we can safely argue, at least in the region of small $y$, effects of the carrier-number change and conduction-electron scattering, separately. Actually, observed characteristics of the electronic specific heat coefficient $\gamma$ of $LaFe_{1-y}Co_yAsO_{0.89}F_{0.11}$ can basically be understood by the rigid band picture,[13)] where Ni atom donates two electrons to the La1111 system and behaves as a nonmagnetic impurity. The rate of the $T_c$ suppression by Ni impurities can be understood by the above idea that the electron number solely determines the $T_c$ values, where the common $p_c$ value is found to those of the Co- and F-doped systems. It confirms that the effect of the pair-breaking by nonmagnetic impurities expected for the $s_\pm$ symmetry of $\Delta$ does not exist. The result strongly supports that the system has the $s_{++}$ symmetry without the sign-reversing of the order parameters.

Experimental details are described in §2. Then, the specific heat data and transport data are presented in §3. 1 and §3. 2, respectively, to show how the rigid band picture describes them, where discussion is also given on whether the pair breaking effect exist or not. In the section, we discuss briefly a randomness effect on the superconductivity induced by the impurity doping, and in §4, the summary is given.

[#] present address: Department of Applied Physics, Tohoku University, Sendai 980-8578, Japan

*present address: Research Center for Neutron Science and Technology, Comprehensive Research Organization for Science and Society, Sirakata, Tokai, Ibaraki 319-1106, Japan

[†]corresponding author: M. Sato (m_sato@cross.or.jp)

## 2. Experiments

Polycrystalline samples of $LaFe_{1-y}Ni_yAsO_{0.89}F_{0.11}$ were prepared from initial mixtures of La, $La_2O_3$, $LaF_3$, FeAs and NiAs at nominal molar ratios. The NiAs powder was obtained by annealing mixtures of Ni- and As-powders in an evacuated quartz tube at 850°C. Detail of the preparations is given in our previous paper.[4] The X-ray powder patterns were obtained with Cu Kα radiation, we have observed an impurity phase LaOF in several samples, whose typical molar concentration is ~0.03. We found that the Ni doping was basically successful, as can be judged from the *y* dependence of the lattice parameters *a* and *c* shown in Fig. 1(a). These results are consistent with those of $LaFe_{1-y}Ni_yAsO$ system.[14, 15] The superconducting diamagnetic moments were measured using a Quantum Design SQUID magnetometer with a magnetic field *H* of 10 G under both zero-field-cooling (ZFC) and field-cooling (FC) conditions. Only the ZFC data are shown in Fig. 1(b). Electrical resistivities were measured by a four-terminal method with an ac-resistance bridge. $T_c$ values were determined as described previously[4] using the data of the superconducting diamagnetic susceptibilities and electrical resistivities ρ. The two kinds of $T_c$ values are in good agreement. Hall coefficients $R_H$ of the polycrystalline samples were measured with a stepwise increase in the temperature *T* at a magnetic field *H* of 7 T, where the sample plates were rotated around the axis perpendicular to the field, and the thermoelectric power *S* was measured by the methods described previously.[16, 17]

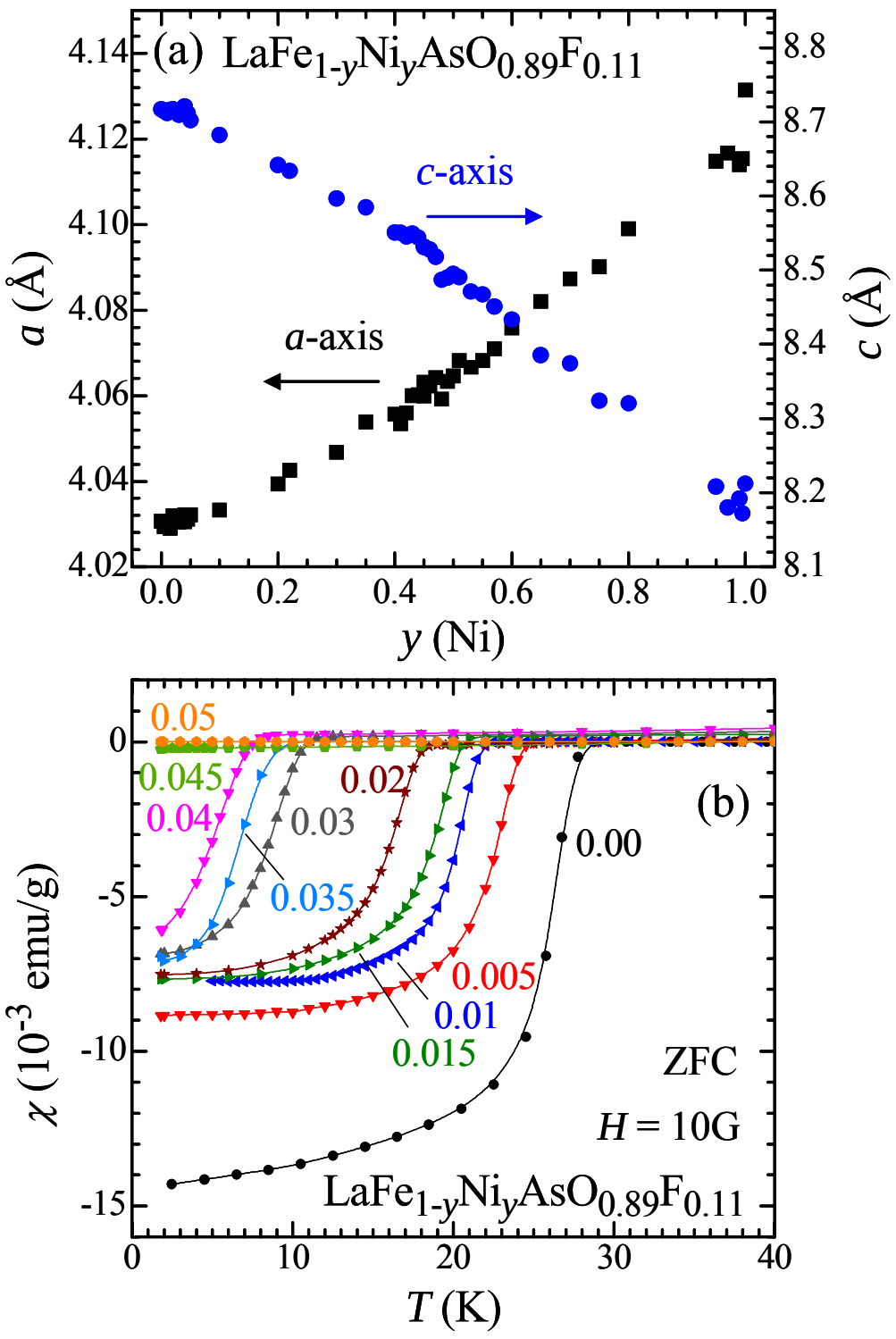

Fig. 1: (color online) (a) Lattice parameters *a* and *c* of $LaFe_{1-y}Ni_yAsO_{0.89}F_{0.11}$ are plotted against *y*. (b) *T* dependence of the magnetic susceptibility obtained under the condition of zero-field-cooling (ZFC) for $LaFe_{1-y}Ni_yAsO_{0.89}F_{0.11}$. The numerical numbers attached to each curve indicate the corresponding *y* value.

## 3. Results and Discussion

### 3.1 Rigid band picture

In Fig. 2, data of the specific heat are shown in the form of $C/T$-$T^2$ for several samples of $LaFe_{1-y}Ni_yAsO_{0.89}F_{0.11}$. Several non-superconducting samples exhibit an upturn with decreasing *T* due to the magnetically active nature of the systems at low temperatures and superconducting samples exhibit a jump at $T_c$. For the non-superconducting samples, the γ values were estimated by fitting the equation $C/T = \gamma + \beta T^2 + \delta T^4$ to the observed data. In this fitting, we neglected the upturn of the specific heat observed at low temperature with decreasing *T*. For the superconducting samples, on the other hand, we estimated the jump magnitudes $\Delta C$ at $T_c$ and calculated the values of γ, simply assuming the relation $\Delta C/T_c = 1.43\gamma$.

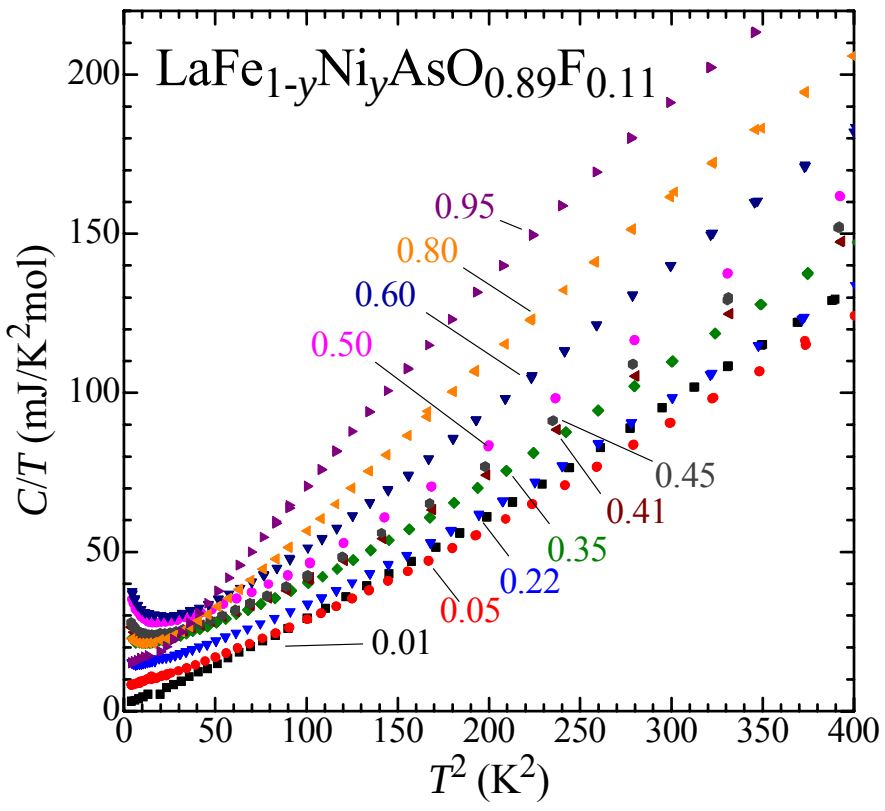

Fig. 2: (color online) Specific heat *C* divided by temperature *T* is plotted against $T^2$ for several samples of $LaFe_{1-y}Ni_yAsO_{0.89}F_{0.11}$. The numerical number attached to each curve indicates the corresponding *y* value.

Figure 3(b) shows the *p* dependence of γ obtained experimentally for $LaFe_{1-y}Ni_yAsO_{0.89}F_{0.11}$ together with those for $LaFe_{1-y}M_yAsO_{0.89}F_{0.11}$ (M = Co, Mn) systems.[6] In this figure, we used equations $p = 2y + 0.11$, $p = y + 0.11$ and $p = -y + 0.11$ for the Ni-, Co- and Mn-doped systems, respectively. Actually, comparing the results of the band calculation for LaFeAsO shown in Fig. 3(a), which show the electronic density of states $N(E)$ against the electron energy *E*,[13] we find that characteristics of the electronic *p* or *E* dependence of γ or $N(E)$ are very similar, although we realize two discrepancies between the observed and calculated results. The first one is that there is no sharp peak at $p$ ~1.0 for the Ni-doped system, It is possibly due to the smearing of $N(E)$ by an effect of severe randomness introduced by Ni impurities. In the Co-doping, on the other hand, because the peak position almost corresponds to the pure Co system [see Fig. 3(a), where the chemical potentials for LaCoAsO and LaNiAsO are shown by the arrows.[13, 18]], it is not smeared out, and a ferromagnetic phase transition originating from the large $N(E)$ value takes place at $T_C$ ~ 60K.[19, 20] The second discrepancy is that in the *y* region of the Ni-doped system, [(2*y*+0.11)>1.3], the observed γ begins to decrease with increasing *p* or *E*, while the calculated $N(E)$ does not. It may be due to the decrease of the electron effective mass, which has actually been observed in the behaviors of the resistivity and magnetic susceptibility, as the system approaches LaNiAsO.

In the present studies, it has been shown how the rigid band picture is applicable to the present system, where Ni impurity can be considered to be nonmagnetic and donate 2 electrons to the host bands.

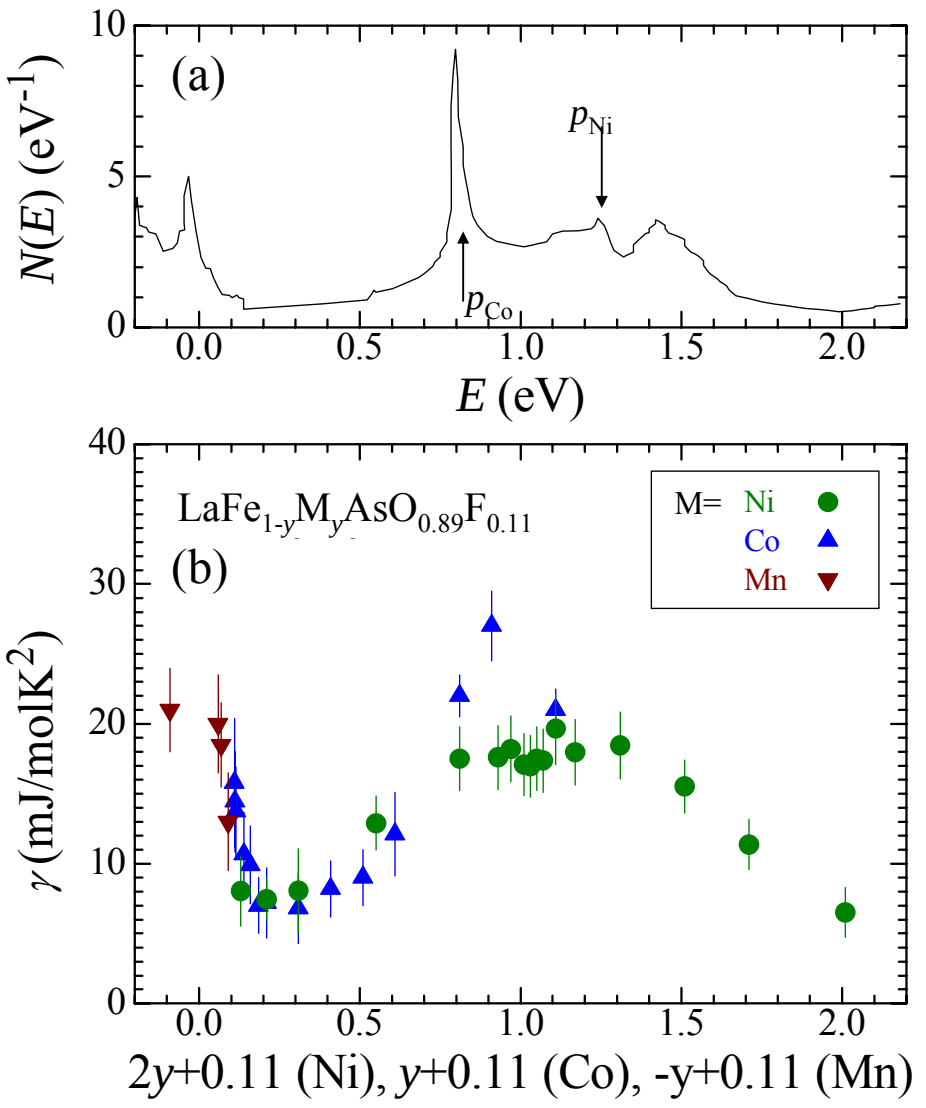

Fig. 3: (color online) (a) Electronic density of states calculated for LaFeAsO against the electron energy.[13)] (b) The electron-number ($p$) dependence of the electronic specific heat coefficient $\gamma$ for LaFe$_{1\text{-}y}$Ni$_y$AsO$_{0.89}$F$_{0.11}$.

### 3.2 Transport properties

Figure 4 shows the temperature dependence of ρ for the several samples of LaFe$_{1\text{-}y}$Ni$_y$AsO$_{0.89}$F$_{0.11}$. The superconducting transition appears in the region of $y \lesssim 0.04$, and samples with $y \gtrsim 0.05$ exhibit non-superconducting metallic behavior. For $y$ = 0, ρ exhibits the S-shape $T$ dependence. This behavior seems to become weaker, as $y$ becomes larger. The anomalous behavior has also been observed in the thermoelectric power $S$ and Hall coefficient $R_H$, as shown in Figs. 5(a) and 5(c), respectively. For samples with $y \le 0.05$ ($p \le 0.2$), both $S$ and $R_H$ exhibit strong $T$ dependence similar to those observed in high-$T_c$ Cu oxides.[16, 17)] These anomalous behaviors have been observed in various Fe pnictide systems.[4-8)] They indicate that the system is magnetically active, similarly to the superconducting Cu oxides.[16, 17, 21)] The observed decrease of $|S|$ and $|R_H|$ with increasing $y$ can be understood by considering the weakening of the magnetic fluctuation due to the reduction of the area of the hole Fermi surface. The decreasing rates, d$|S|$/d$y$ and d$|R_H|$/d$y$ observed in relatively low $T$ region (but above $T_c$) for the Ni-doped system are roughly two times larger than those for the Co-doped system.[6)] This result also supports the idea that Co- and Ni-atoms donate one and two electrons to the conduction band, respectively. Figure 6 shows the $y$ dependence of $\rho_0$ for LaFe$_{1\text{-}y}$Ni$_y$AsO$_{0.89}$F$_{0.11}$ together with that for LaFe$_{1\text{-}y}$Co$_y$AsO$_{0.89}$F$_{0.11}$.[6)] The values of $\rho_0$ were estimated by the extrapolation of the ρ-$T$ curve to $T$ = 0 from the $T$ region where the resistivity upturn with decreasing $T$ is not significant. The initial value of d$\rho_0$/d$y$ is nearly equal to that for the Co-doped system,[6)] indicating that the scattering strength of Ni impurities is similar to that of Co atoms. In Fig. 7, the $T_c$ values of the Ni-doped system are shown against the $x$+2$y$+0.11. In the figure, those of Co-, Ru- and F-doped systems[6, 8)] are also plotted against the $x$+$y$+0.11. (In these figures, $x$=0 for Co-, Ni- and Ru-doped system, and $y$=0 for F-doped system). For Ni-, Co- and F-doped systems, the horizontal axis indicates the electron numbers $p$ doped into LaFeAsO. For the Ru-doped system, on the other hand, the horizontal axis just indicates the Ru atom concentration (note that the electron number does not change with $y$). In the figure, the initial slope of the curve shown for the Ni-doped system is almost equal to those for the Co- and F-doped systems and $T_c$ disappears at the critical value $p_c \sim 0.2$, which corresponds to the disappearance of the hole Fermi surface. For the Ru-doped system, the initial absolute slope is very small, which supports that $T_c$ basically depends only on the carrier number.

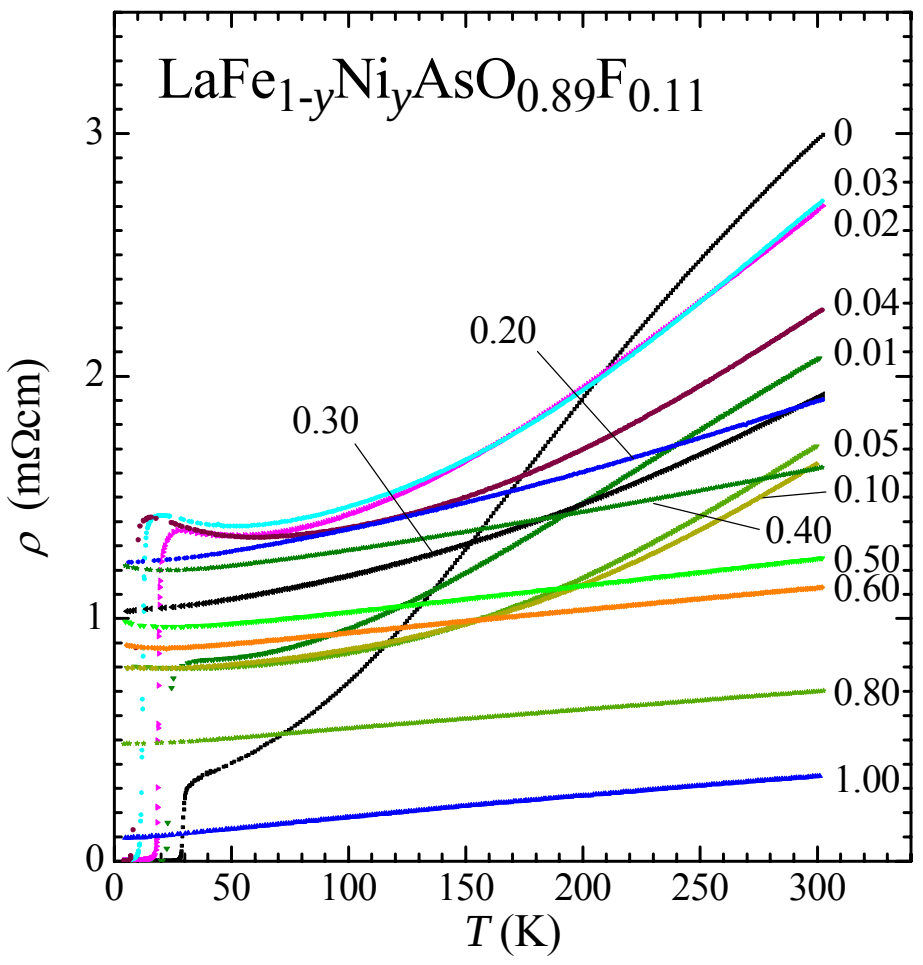

Fig. 4: (color online) $T$ dependence of the resistivities ρ of LaFe$_{1\text{-}y}$Ni$_y$AsO$_{0.89}$F$_{0.11}$. The numerical number attached to each curve indicates the corresponding $y$ value.

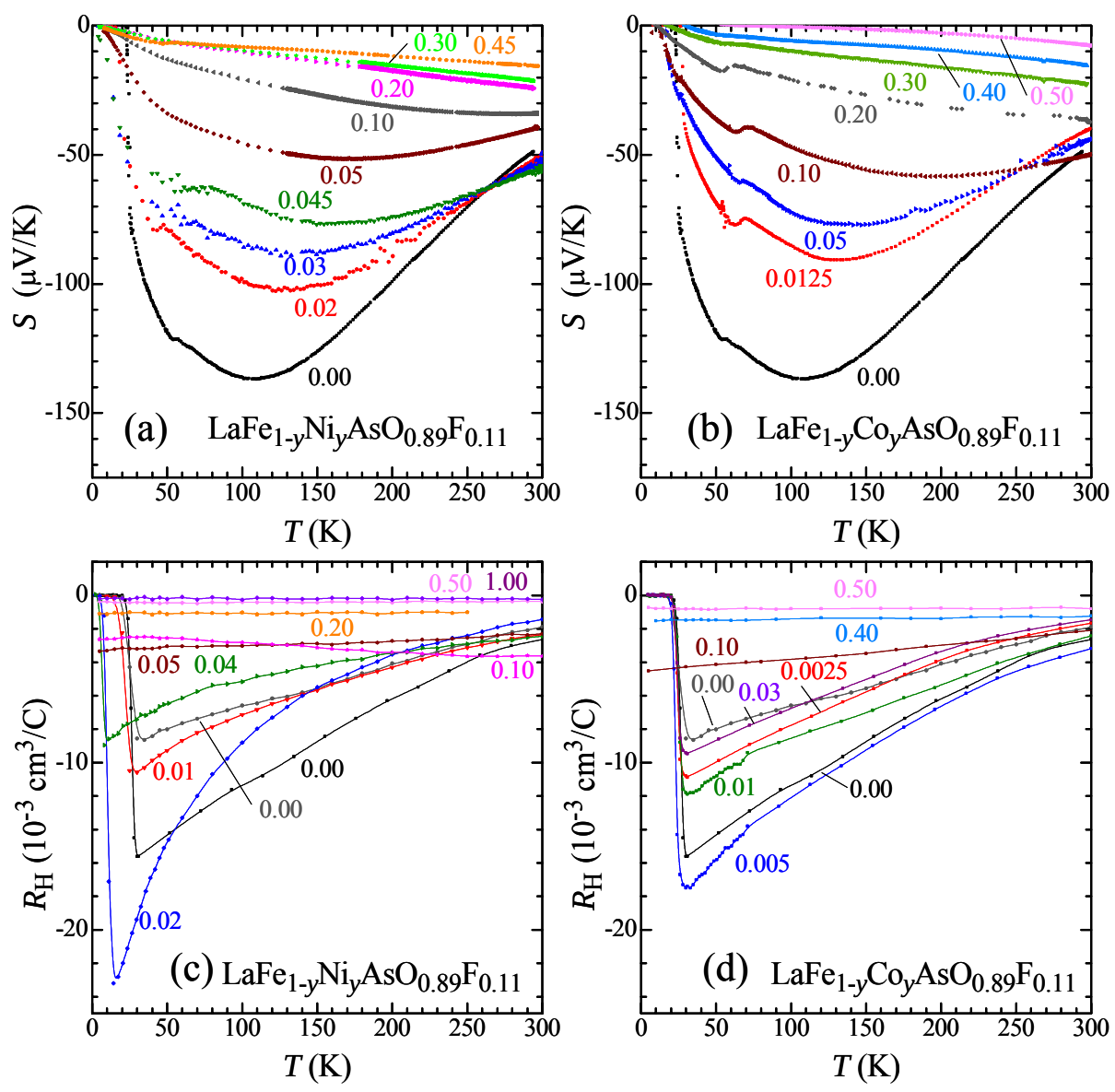

Fig. 5: (color online) $T$ dependences of the thermoelectric power $S$ of (a) LaFe$_{1\text{-}y}$Ni$_y$AsO$_{0.89}$F$_{0.11}$ and (b) LaFe$_{1\text{-}y}$Co$_y$AsO$_{0.89}$F$_{0.11}$.[6)] $T$ dependences of Hall coefficient $R_H$ of (c) LaFe$_{1\text{-}y}$Ni$_y$AsO$_{0.89}$F$_{0.11}$ and (d) LaFe$_{1\text{-}y}$Co$_y$AsO$_{0.89}$F$_{0.11}$.[6)] The numerical number attached to each curve indicates the corresponding $y$ value.

If superconductors have the $s_\pm$ symmetry, the initial rate of the $T_c$ suppression by nonmagnetic impurities can be estimated by a simple pair breaking equation,

$$\ln\left(\frac{T_{c0}}{T_c}\right) = \psi\left(\frac{1}{2}+\frac{\alpha}{2t}\right) - \psi\left(\frac{1}{2}\right)$$

where ψ(z) is the digamma function defined as ψ(z) ≡ ln[$d\Gamma(z)/dz/\Gamma(z)$], $T_{c0}$ is the superconducting $T_c$ of the nondoped system, and $t$ = $T_c/T_{c0}$. The pair breaking parameter $\alpha \equiv \hbar/(2\pi k_B T_{c0}\tau)$ can be calculated from the initial rate of $\rho_0$ and the carrier number estimated from the value of $R_H$ by a method described in previous paper in detail.[6)] For the present system, $T_c$ has to disappear at a $y$ value smaller than 0.0035, which should be compared with the observed value of ~0.04-0.05. This argument indicates that not only for Co- and Ru-doped systems but also for the Ni-doped system, the observed $T_c$-suppression rates cannot be understood by considering the pair-breaking effect, that is, the system does not have the $s_\pm$ symmetry of the order parameter.

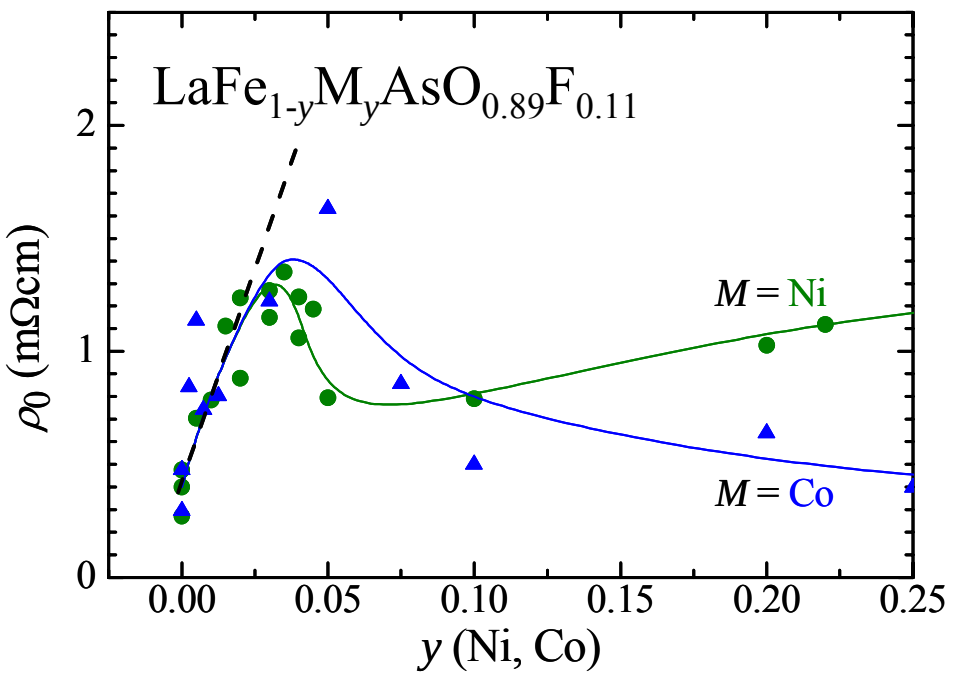

Fig. 6: (color online) Residual resistivities $\rho_0$ of $LaFe_{1-y}Ni_yAsO_{0.89}F_{0.11}$ and $LaFe_{1-y}Co_yAsO_{0.89}F_{0.11}$[6] are plotted against $y$. The broken line shows an initial rate of the $\rho_0$ increase with $y$.

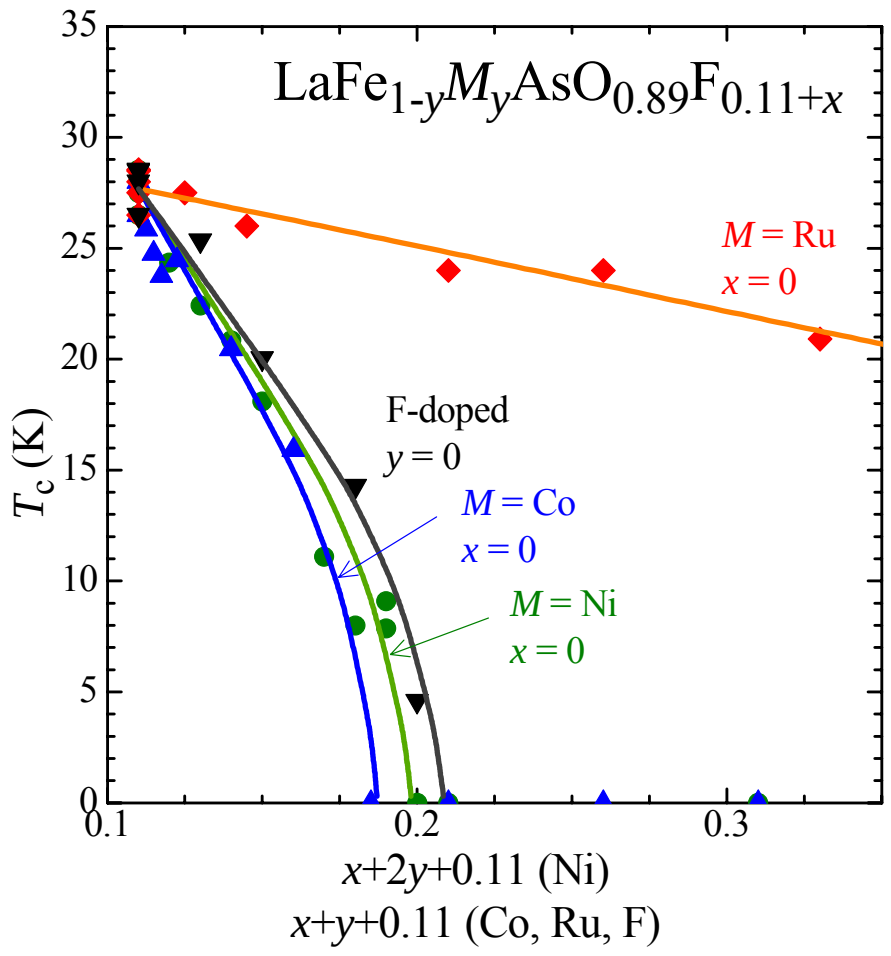

Fig. 7: (color online) $T_c$ values determined from the resistivity measurements are plotted for $LaFe_{1-y}Ni_yAsO_{0.89}F_{0.11}$ against $x+2y+0.11$ and for $LaM_yAsO_{0.89-x}F_{0.11+x}$ against $x+y+0.11$ ($x$ = 0 for M = Co[6] and Ru,[8]) and $y$ = 0 for the sample with nonzero $x$.[8]).

Here, we present a brief comment on the relatively large decrease of $T_c$ reported by Guo *et al.* for Zn-doped La1111 system.[22] As we discussed in detail previously,[6] their results obtained for samples prepared under high pressure can be naturally explained by the well-known effect of the electron localization: When the sheet resistance $R_\square$ defined by using the lattice parameter $c$ as $R_\square \equiv \rho_0/c$ exceeds $h/4e^2$~6.45 kΩ, $T_c$ disappears. It is also stressed that for the present multi band system, the absolute slope $|dT_c/dy|$ due to the electron scattering by nonmagnetic impurities is significantly larger, if it has the $s_\pm$ symmetry, than that expected for single band Cu oxide superconductors, even though the latter systems also have the sign-reversing Δ. It is because impurities such as Zn atoms doped into the conducting planes of the single band superconductors act as the unitary scatterers, for which the $T_c$ suppression effect is much reduced.[23] Therefore, for Fe pnictide systems, even if observed $|dT_c/dy|$ is as large as that observed for Zn–doped Cu oxide superconductors, it does not indicate the existence of the pair breaking effect.

As we reported previously, $|dT_c/dy|$ observed for $LaFe_{1-y}Mn_yAsO_{0.89}F_{0.11}$ is also rather large.[6] However, for the system, the resistivity increases very rapidly with increasing $y$, indicating that the electron localization takes place at the very small value of $y$, and the system rapidly exhibits the magnetic nature. We cannot distinguish which of the magnetism and electron localization is primarily important for the $T_c$ suppression.

## 4. Summary

We have carried out measurements of the specific heats and transport properties of $LaFe_{1-y}Ni_yAsO_{0.89}F_{0.11}$ and examined if the rigid band picture can be applied to the Fe pnictide superconductors. We have also studied the symmetry of the superconducting order parameter. We have found that the characteristics of the electric specific coefficients γ of the Ni- and Co-doped[6] systems can be understood, (at least in the $y$ region described by $[(2y+0.11)<1.3]$), by the band calculation reported for LaFeAsO system,[13] showing the validity of the rigid band picture. On the basis of this result, we can consider that a Ni dopant donates two electrons to this system and also behaves as a nonmagnetic impurity. The $T_c$ suppression can be understood by an idea that the carrier number solely determines the $T_c$ values. No effect of the pair-breaking due to the conduction electron scattering by nonmagnetic impurities have been observed. The results strongly exclude the $s_\pm$ symmetry of the order parameter.

## Acknowledgment

The authors thank Prof. H. Kontani for fruitful discussion. The work is supported by Grants-in-Aid for Scientific Research from the Japan Society for the Promotion of Science (JSPS) and Technology and JST, TRIP.